\documentclass[12pt]{article}
\usepackage{amsmath,amsthm}
\usepackage{eucal}
\usepackage{amsfonts}
\newcommand{\ar}{\renewcommand{\arraystretch}{1}} 
\gdef\C{\Bbb C}
\gdef\dZ{\Bbb Z}

\gdef\dS{\Bbb S}

\DeclareMathOperator{\spin}{{\bf Spin}}

\DeclareMathOperator{\Sym}{Sym}
\DeclareMathOperator{\diver}{div}
\DeclareMathOperator{\rot}{rot}

\newcommand{\cA}{\mathcal{A}}

\newcommand{\cL}{\mathcal{L}}

\newcommand{\sI}{{\sf I}}

\newcommand{\sT}{{\sf T}}

\newcommand{\bj}{{\bf j}}

\newcommand{\bx}{{\bf x}}
\newcommand{\bB}{{\bf B}}

\newcommand{\bF}{{\bf F}}
\newcommand{\bE}{{\bf E}}

\newcommand{\fC}{\mathfrak{C}}

\newcommand{\fG}{\mathfrak{G}}

\newcommand{\cl}{C\kern -0.2em \ell}

\newcommand{\e}{\mbox{\bf e}}

\newcommand{\ld}{\left[}
\newcommand{\rd}{\right]}

\begin{document}
\title{A Note on the Majorana-Oppenheimer Quantum Electrodynamics}
\author{V.~V. Varlamov}
\date{}
\maketitle
\begin{abstract}
A group theoretical description of the Majorana-Oppenheimer quantum
electrodynamics is considered. Different spinor realizations of the
Maxwell and Dirac fields are discussed. A representation of the
Majorana-Oppenheimer wave equations in terms of the Gel'fand-Yaglom
formalism is given.
\end{abstract}
\leftline{{\bf PACS 1998:} 02.10.Tq, 02.30.Gp}
\leftline{{\bf MSC 2000:} 15A66, 22E43, 35Q40}
\vspace{0.5cm}
In 1950, Gupta \cite{Gup50} and Bleuler \cite{Ble50} proposed to quantize
the electromagnetic field via the four-potential $A_\mu$. As is known,
such a quantization gives rise to the following difficulties:
non-physical degrees of freedom, indefinite metric, null helicity, peculiar
opposition between electromagnetic field and other physical fields. More
recently, Bogoliubov and Shirkov said: ``Among all the physical fields the
electromagnetic field is quantized with the most difficulty'' \cite{BS93}.
For that reason at present time many physicists considered a situation
with the quantization of the electromagnetic field as unsatisfactory.
It is clear that the Gupta-Bleuler phenomenology should be replaced by a more
rigorous alternative.

Beyond all shadow of doubt, one of the main candidates on the role of such
an alternative is the Majorana-Oppenheimer quantum electrodynamics.
At the beginning of thirties of the last century
Majorana \cite{Maj} and Oppenheimer \cite{Opp31} proposed to consider
the Maxwell theory of electromagnetism as the wave mechanics of the photon.
They introduced a wave function of the form
\begin{equation}\label{MO1}
\boldsymbol{\psi}=\bE-i\bB,
\end{equation}
where $\bE$ and $\bB$ are electric and magnetic fields. In virtue of this
the standard Maxwell equations can be rewritten
\begin{equation}\label{1}
\diver\boldsymbol{\psi}=\rho,\quad i\rot\boldsymbol{\psi}=\bj+
\frac{\partial\boldsymbol{\psi}}{\partial t},
\end{equation}
where $\boldsymbol{\psi}=(\psi_1,\psi_2,\psi_3)$, $\psi_k=E_k-iB_k$
($k=1,2,3$). In accordance with correspondence principle
$\left(-i\partial/\partial x_i\rightarrow p_i; +i\partial/\partial t
\rightarrow W\right)$ and in absence of electric charges and currents
the equations (\ref{1}) take a Dirac-like form
\begin{equation}\label{2}
(W-\boldsymbol{\alpha}\cdot\boldsymbol{p})\boldsymbol{\psi}=0
\end{equation}
with transversality condition
\[
\boldsymbol{p}\cdot\boldsymbol{\psi}=0.
\]
At this point, three matrices
\begin{equation}\label{3}
\alpha^1=\begin{pmatrix}
0 & 0 & 0\\
0 & 0 & i\\
0 &-i & 0
\end{pmatrix},\quad\alpha^2=\begin{pmatrix}
0 & 0 & -i\\
0 & 0 & 0\\
i & 0 & 0
\end{pmatrix},\quad\alpha^3=\begin{pmatrix}
0 & i & 0\\
-i& 0 & 0\\
0 & 0 & 0
\end{pmatrix}
\end{equation}
satisfy the angular-momentum commutation rules
\[
\ld\alpha_i,\alpha_k\rd=-i\varepsilon_{ikl}\alpha_l\quad
(i,k,l=1,2,3).
\]
Further, for the complex conjugate wave function
\begin{equation}\label{MO2}
\boldsymbol{\psi}^\ast=\bE+i\bB
\end{equation}
there are the analogous Dirac-like equations
\begin{equation}\label{4}
(W+\boldsymbol{\alpha}\cdot\boldsymbol{p})\boldsymbol{\psi}^\ast=0.
\end{equation}
From the equations (\ref{2}) and (\ref{4}) it follows that photons coincide
with antiphotons (truly neutral particles). In such a way,
$\boldsymbol{\psi}$ ($\boldsymbol{\psi}^\ast$) may be considered as
a wave function of the photon satisfying the massless Dirac-like equations.
In contrast to the Gupta-Bleuler phenomenology, where the
non-observable four-potential $A_\mu$ is quantized, the main
advantage of the Majorana-Oppenheimer formulation of electrodynamics
lies in the fact that it deals directly with observable quantities,
such as the electric and magnetic fields.

Let us consider now a relationship between the Majorana-Oppenheimer
formulation of electrodynamics and a group theoretical framework of
quantum field theory. It is widely accepted that the Lorentz group
(a rotation group of the 4-dimensional space-time continuum) is a kernel
of relativistic physics. Fields and particles are completely formulated
within irreducible representations of the Lorentz or Poincar\'{e} group
(see Wigner and Weinberg works \cite{Wig39,Wein}, where physical fields
are considered in terms of induced representations of the Poincar\'{e}
group). According to Wigner \cite{Wig39}, a quantum system, described
by an irreducible representation of the Poincar\'{e} group, is called
an elementary particle. The group theoretical formulation of quantum field
theory allows one to describe all the physical fields in equal footing,
without any division on `gauge' and `matter' fields as it accepted in
modern gauge phenomenologies, such as Standard model and so on.
On the other hand, there is a close relationship between linear
representations of the Lorentz group and Clifford algebras (for more
details see \cite{Var01,Var02a} and references therein).
As is known, a double covering of the proper orthochronous Lorentz group $\fG_+$,
the group $SL(2,\C)$, is isomorphic to the Clifford--Lipschitz group
$\spin_+(1,3)$, which, in its turn, is fully defined within a
biquaternion algebra $\C_2$, since
\[\ar
\spin_+(1,3)\simeq\left\{\begin{pmatrix} \alpha & \beta \\ \gamma & \delta
\end{pmatrix}\in\C_2:\;\;\det\begin{pmatrix}\alpha & \beta \\ \gamma & \delta
\end{pmatrix}=1\right\}=SL(2,\C).
\]\begin{sloppypar}\noindent
Thus, a fundamental representation of the group $\fG_+$ is realized in a
spinspace $\dS_2$. The spinspace $\dS_2$ is a complexification of the
minimal left ideal of the algebra $\C_2$: $\dS_2=\C\otimes I_{2,0}=\C\otimes
\cl_{2,0}e_{20}$ or $\dS_2=\C\otimes I_{1,1}=\C\otimes\cl_{1,1}e_{11}$
($\C\otimes I_{0,2}=\C\otimes\cl_{0,2}e_{02}$), where $\cl_{p,q}$
($p+q=2$) is a real subalgebra of $\C_2$, $I_{p,q}$ is the minimal left ideal
of the algebra $\cl_{p,q}$, $e_{pq}$ is a primitive idempotent.
\end{sloppypar}
Further, let $\overset{\ast}{\C}_2$ be the biquaternion algebra, in which
all the coefficients are complex conjugate to the coefficients of the
algebra $\C_2$. The algebra $\overset{\ast}{\C}_2$ is obtained from
$\C_2$ under action of the automorphism $\cA\rightarrow\cA^\star$
(involution), or the antiautomorphism $\cA\rightarrow\widetilde{\cA}$ (reversal),
where $\cA\in\C_2$ (see \cite{Var99,Var00,Var03}). Let us compose a tensor
product of $k$ algebras $\C_2$ and $r$ algebras $\overset{\ast}{\C}_2$:
\begin{equation}\label{Ten}
\C_2\otimes\C_2\otimes\cdots\otimes\C_2\otimes\overset{\ast}{\C}_2\otimes
\overset{\ast}{\C}_2\otimes\cdots\otimes\overset{\ast}{\C}_2\simeq
\C_{2k}\otimes\overset{\ast}{\C}_{2r}.
\end{equation}
The tensor product (\ref{Ten}) induces a spinspace
\begin{equation}\label{Spin}
\dS_2\otimes\dS_2\otimes\cdots\otimes\dS_2\otimes\dot{\dS}_2\otimes
\dot{\dS}_2\otimes\cdots\otimes\dot{\dS}_2=\dS_{2^{k+r}}
\end{equation}
with `vectors' (spintensors) of the form
\begin{equation}\label{Vect}
\xi^{\alpha_1\alpha_2\cdots\alpha_k\dot{\alpha}_1\dot{\alpha}_2\cdots
\dot{\alpha}_r}=\sum\xi^{\alpha_1}\otimes\xi^{\alpha_2}\otimes\cdots\otimes
\xi^{\alpha_k}\otimes\xi^{\dot{\alpha}_1}\otimes\xi^{\dot{\alpha}_2}\otimes
\cdots\otimes\xi^{\dot{\alpha}_r}.
\end{equation}\begin{sloppypar}\noindent
The full representation space $\dS_{2^{k+r}}$ contains both symmetric and
antisymmetric spintensors (\ref{Vect}). Usually, at the definition of
irreducible finite-dimensional representations of the Lorentz group
physicists confined to a subspace of symmetric spintensors
$\Sym(k,r)\subset\dS_{2^{k+r}}$. Dimension of $\Sym(k,r)$ is equal to
$(k+1)(r+1)$ or $(2l+1)(2l^\prime+1)$ at $l=\frac{k}{2}$,
$l^\prime=\frac{r}{2}$.
The space
$\Sym(k,r)$ can be considered as a space of polynomials\end{sloppypar}
\begin{gather}
p(z_0,z_1,\bar{z}_0,\bar{z}_1)=\sum_{\substack{(\alpha_1,\ldots,\alpha_k)\\
(\dot{\alpha}_1,\ldots,\dot{\alpha}_r)}}\frac{1}{k!\,r!}
a^{\alpha_1\cdots\alpha_k\dot{\alpha}_1\cdots\dot{\alpha}_r}
z_{\alpha_1}\cdots z_{\alpha_k}\bar{z}_{\dot{\alpha}_1}\cdots
\bar{z}_{\dot{\alpha}_r}\label{SF}\\
(\alpha_i,\dot{\alpha}_i=0,1),\nonumber
\end{gather}
where the numbers
$a^{\alpha_1\cdots\alpha_k\dot{\alpha}_1\cdots\dot{\alpha}_r}$
are unaffected at the permutations of indices. Some applications of the
functions (\ref{SF}) contained in \cite{Vas96,GS00}.

In accordance with the Wigner-Weinberg scheme all the physical fields
are described within $\boldsymbol{\tau}_{l\dot{l}}$-representations of
the proper Lorentz group $\fG_+=SL(2,\C)/\dZ_2$ or within the proper
Poincar\'{e} group $SL(2,\C)\odot T_4/\dZ_2$, where $T_4$ is a subgroup
of four-dimensional translations (since $T_4$ is an Abelian group, then
all its representations are one-dimensional). At this point, any
representation $\boldsymbol{\tau}_{l\dot{l}}$ can be written as
$\boldsymbol{\tau}_{l0}\otimes\boldsymbol{\tau}_{0\dot{l}}$, where
$\boldsymbol{\tau}_{l0}$ ($\boldsymbol{\tau}_{0\dot{l}}$) are representations
of the group $SU(2)$. In a sence, it allows us to represent the group
$SL(2,\C)$ by a product $SU(2)\otimes SU(2)$ as it done by Ryder in his
textbook \cite{Ryd85}. Moreover, in the works \cite{Ahl93,Dvo96} the
Lorentz group is represented by a product $SU_R(2)\otimes SU_L(2)$,
where the spinors $\psi(p^\mu)=\begin{pmatrix}\phi_R(p^\mu)\\
\phi_L(p^\mu)\end{pmatrix}$ ($\phi_R(p^\mu)$ and $\phi_L(p^\mu)$ are
the right- and left-handed spinors) are transformed within
$(j,0)\oplus(0,j)$ representation space, in our case $j=l=\dot{l}$.
All these representations for $SL(2,\C)$ follow from the Van der Waerden
representation of the Lorentz group which was firstly given in the
brilliant book \cite{Wa32} and further widely accepted by many authors
\cite{AB,Sch61,RF,Ryd85}. In \cite{Wa32} the group $SL(2,\C)$ is
understood as a complexification of the group $SU(2)$,
$SL(2,\C)\sim\mbox{\sf complex}(SU(2))$ (see also \cite{Var02b}).
Further, in accordance with the Wigner-Weinberg scheme a helicity
$\lambda$ of the particle, described by the representations
$\boldsymbol{\tau}_{l\dot{l}}$, is defined by an expression
$l-\dot{l}=\lambda$ (Weinberg Theorem \cite{Wein}). However,
the Wigner-Mackey-Weinberg scheme does not incorporate the Clifford
algebraic framework of the Lorentz group. On the other hand, there is
a more powerful method based on a generalized regular representation
\cite{Vil68,VK} (GGR-theory) which naturally incorporates Clifford
algebraic description \cite{Var02b} and includes the Wigner-Weinberg
approach as a particular case.

As is known, basic equations of quantum electrodynamics in any formulation
are the Maxwell equations
\begin{eqnarray}
\diver\bE&=&4\pi\rho,\nonumber\\
\diver\bB&=&0,\nonumber\\
\rot\bB&=&\frac{4\pi}{c}\bj+\frac{1}{c}\frac{\partial\bE}{\partial t},
\nonumber\\
\rot\bE&=&-\frac{1}{c}\frac{\partial\bB}{\partial t}\label{Maxwell}
\end{eqnarray}
and the Dirac equations
\begin{equation}\label{Dirac}
\left(i\gamma_\mu\frac{\partial}{\partial x_\mu}-m\right)\psi(x)=0,
\end{equation}
where $\gamma$-matrices in a canonical basis have the form
\[
\gamma_0=\begin{pmatrix}
\sigma_0 & 0\\
0 & \sigma_0
\end{pmatrix},\;\;\gamma_1=\begin{pmatrix}
0 & \sigma_1\\
-\sigma_1 & 0
\end{pmatrix},\;\;\gamma_2=\begin{pmatrix}
0 & \sigma_2\\
-\sigma_2 & 0
\end{pmatrix},\;\;\gamma_3=\begin{pmatrix}
0 & \sigma_3\\
-\sigma_3 & 0
\end{pmatrix}.
\]
It is well known that Maxwell and Dirac equations can be rewritten in
a spinor form (see \cite{LU31,Rum36,RF}). In this form these equations
look very similar. Indeed, the spinor form of the equations (\ref{Maxwell})
and (\ref{Dirac}) is
\[
\begin{array}{ccc}
\partial^{\dot{\mu}}_\lambda f^\lambda_\rho&=&0,\\
\partial^{\dot{\mu}}_\lambda f^\lambda_\rho&=&s^{\dot{\mu}}_\rho;
\end{array}\quad\quad\begin{array}{ccc}
\partial^{\lambda\dot{\mu}}\eta_{\dot{\mu}}+
im\xi^\lambda&=&0,\\
\partial_{\lambda\dot{\mu}}\xi^\lambda+
im\eta_{\dot{\mu}}&=&0.
\end{array}
\]
Moreover, in the vacuum ($s^{\dot{\mu}}_\rho=0$) and for the massless
field ($m=0$) these equations take the form
\[
\begin{array}{ccc}
\partial_{1\dot{1}}f_{11}+\partial_{1\dot{2}}f_{12}&=&0,\\
\partial_{2\dot{1}}f_{11}+\partial_{2\dot{2}}f_{12}&=&0,\\
\partial^{1\dot{1}}f^{\dot{1}\dot{1}}+
\partial^{2\dot{1}}f^{\dot{1}\dot{2}}&=&0,\\
\partial^{1\dot{2}}f^{\dot{1}\dot{1}}+
\partial^{2\dot{1}}f^{\dot{1}\dot{2}}&=&0,
\end{array}\quad\begin{array}{ccc}
\partial_{1\dot{1}}\xi^1+\partial_{1\dot{2}}\xi^2&=&0,\\
\partial_{2\dot{1}}\xi^1+\partial_{2\dot{2}}\xi^2&=&0,\\
\partial^{1\dot{1}}\eta_{\dot{1}}+
\partial^{2\dot{1}}\eta_{\dot{2}}&=&0,\\
\partial^{1\dot{2}}\eta_{\dot{1}}+
\partial^{2\dot{1}}\eta_{\dot{2}}&=&0.
\end{array}
\]
The latter equations formally coincide if we suppose
$\xi^\lambda=(f_{11},f_{12})^{\sT}$,
$\eta_{\dot{\mu}}=(f^{\dot{1}\dot{1}},f^{\dot{1}\dot{2}})^{\sT}$ as it done
by Rumer in \cite{Rum36}.
In spite of this similarity there is a deep difference
between Dirac and Maxwell equations. Namely, spinors $\xi^\lambda$ and
$\eta_{\dot{\mu}}$ are transformed within
$\boldsymbol{\tau}_{\frac{1}{2},0}$ and
$\boldsymbol{\tau}_{0,\frac{1}{2}}$ representations, whereas the
spintensors $f^{\lambda\mu}$ and $f^{\dot{\lambda}\dot{\mu}}$ are
transformed within $\boldsymbol{\tau}_{1,0}$ and
$\boldsymbol{\tau}_{0,1}$ representations of the Lorentz group. Indeed,
the Dirac electron-positron field
$(1/2,0)\oplus(0,1/2)$ corresponds to the algebra $\C_2\oplus
\overset{\ast}{\C}_2$. It should be noted that the Dirac algebra
$\C_4$, considered as a tensor product $\C_2\otimes\C_2$
(or $\C_2\otimes\overset{\ast}{\C}_2$),
leads to spintensors $\xi^{\alpha_1\alpha_2}$
(or $\xi^{\alpha_1\dot{\alpha}_1}$), but it contradicts with the usual
definition of the Dirac bispinor as a pair
$(\xi^{\alpha_1},\xi^{\dot{\alpha}_1})$. Therefore, the Clifford algebra
associated with the Dirac field is $\C_2\oplus\overset{\ast}{\C}_2$, and
a spinspace of this sum in virtue of unique decomposition
$\dS_2\oplus\dot{\dS}_2=\dS_4$ ($\dS_4$ is a spinspace of $\C_4$) allows one to
define $\gamma$-matrices in the Weyl basis.

In contrast to the Dirac field $(1/2,0)\oplus(0,1/2)$, the Maxwell field
is defined in terms of spintensors of the second rank. According to
(\ref{Ten}), the Clifford algebras, corresponded to the Maxwell fields, are
$\C_2\otimes\C_2$ and
$\overset{\ast}{\C}_2\otimes\overset{\ast}{\C}_2$\footnote{On the other hand,
these algebras can be obtained
by means of an homomorphic mapping
$\epsilon:\,\C_5\rightarrow\C_4$ (see \cite{Var01b}). Indeed,
for the algebra $\C_5$ we have a decomposition
\[
\unitlength=0.5mm
\begin{picture}(70,50)
\put(35,40){\vector(2,-3){15}}
\put(35,40){\vector(-2,-3){15}}
\put(32.25,42){$\C_{5}$}
\put(16,28){$\lambda_{+}$}
\put(49.5,28){$\lambda_{-}$}
\put(13.5,9.20){$\C_{4}$}
\put(52.75,9){$\stackrel{\ast}{\C}_{4}$}
\put(32.5,10){$\cup$}
\end{picture}
\]
where the central idempotents
$
\lambda_+=\frac{1+\e_1\e_2\e_3\e_4\e_5}{2},\;
\lambda_-=\frac{1-\e_1\e_2\e_3\e_4\e_5}{2}
$
correspond to the helicity projection operators of the Maxwell field.
As is known, for the photon there are two helicity states: left and right
handed polarizations. Hence it follows that
in common with other massless fields (such as the neutrino field
$(1/2,0)\cup(0,1/2)$) the Maxwell electromagnetic field is also described
within the quotient representations of the Lorentz group \cite{Var01a}.
In accordance
with Theorem 4 in \cite{Var01} the photon field can be described by a
quotient representation of the class
${}^\chi\fC^{2,0}_{a_1}\cup{}^\chi\fC^{0,-2}_{a_1}$. This representation
admits time reversal $T$ and an identical charge conjugation $C\sim\sI$
that corresponds to truly neutral particles (see Theorem 3 in \cite{Var01}).}
The algebras $\C_2\otimes\C_2$ and
$\overset{\ast}{\C}_2\otimes\overset{\ast}{\C}_2$ induce spinspaces
$\dS_2\otimes\dS_2$ and $\dot{\dS}_2\otimes\dot{\dS}_2$.
These spinspaces are full representation spaces for tensor products
$\boldsymbol{\tau}_{\frac{1}{2},0}\otimes
\boldsymbol{\tau}_{\frac{1}{2},0}$ and
$\boldsymbol{\tau}_{0,\frac{1}{2}}\otimes
\boldsymbol{\tau}_{0,\frac{1}{2}}$. In accordance with (\ref{Vect}) the
basis `vectors' (spintensors) of the spinspaces $\dS_2\otimes\dS_2$ and
$\dot{\dS}_2\otimes\dot{\dS}_2$ have the form
\begin{eqnarray}
&&\xi^{11}=\xi^1\otimes\xi^1,\;\;\xi^{12}=\xi^1\otimes\xi^2,\;\;
\xi^{21}=\xi^2\otimes\xi^1,\;\;\xi^{22}=\xi^2\otimes\xi^2,\nonumber\\
&&\xi^{\dot{1}\dot{1}}=\xi^{\dot{1}}\otimes\xi^{\dot{1}},\;\;
\xi^{\dot{1}\dot{2}}=\xi^{\dot{1}}\otimes\xi^{\dot{2}},\;\;
\xi^{\dot{2}\dot{1}}=\xi^{\dot{2}}\otimes\xi^{\dot{1}},\;\;
\xi^{\dot{2}\dot{2}}=\xi^{\dot{2}}\otimes\xi^{\dot{2}}.\label{ST}
\end{eqnarray}
Further, the representations
$\boldsymbol{\tau}_{\frac{1}{2},0}\otimes\boldsymbol{\tau}_{\frac{1}{2},0}$ and
$\boldsymbol{\tau}_{0,\frac{1}{2}}\otimes\boldsymbol{\tau}_{0,\frac{1}{2}}$
are reducible. In virtue of the Clebsh--Gordan formula
\[
\boldsymbol{\tau}_{l_1,\dot{l}_1}\otimes\boldsymbol{\tau}_{l_2,\dot{l}_2}=
\sum_{|l_1-l_2|\leq k\leq l_1+l_2;|\dot{l}_1-\dot{l}_2|\leq\dot{k}\leq
\dot{l}_1+\dot{l}_2}\boldsymbol{\tau}_{k,\dot{k}}
\]
we have
\begin{eqnarray}
\boldsymbol{\tau}_{\frac{1}{2},0}\otimes\boldsymbol{\tau}_{\frac{1}{2},0}
&=&\boldsymbol{\tau}_{0,0}\oplus\boldsymbol{\tau}_{1,0},\nonumber\\
\boldsymbol{\tau}_{0,\frac{1}{2}}\otimes\boldsymbol{\tau}_{0,\frac{1}{2}}
&=&\boldsymbol{\tau}_{0,0}\oplus\boldsymbol{\tau}_{0,1}.\nonumber
\end{eqnarray}
At this point, the spinspaces $\dS_2\otimes\dS_2$ and
$\dot{\dS}_2\otimes\dot{\dS}_2$ decompose into direct sums of the
symmetric representation spaces:
\begin{eqnarray}
\dS_2\otimes\dS_2&=&\Sym_{(0,0)}\oplus\Sym_{(2,0)},\nonumber\\
\dot{\dS}_2\otimes\dot{\dS}_2&=&\Sym_{(0,0)}\oplus\Sym_{(0,2)}.\nonumber
\end{eqnarray}
The spintensors $\xi^{11}$, $\xi^{12}=\xi^{21}$, $\xi^{22}$ and
$\xi^{\dot{1}\dot{1}}$, $\xi^{\dot{1}\dot{2}}=\xi^{\dot{2}\dot{1}}$,
$\xi^{\dot{2}\dot{2}}$, obtained after symmetrization from (\ref{ST})
compose the bases of three-dimensional complex spaces $\Sym_{(2,0)}$ and
$\Sym_{(0,2)}$, respectively. Let us introduce independent complex
coordinates $F_1$, $F_2$, $F_3$ and $\overset{\ast}{F}_1$,
$\overset{\ast}{F}_2$, $\overset{\ast}{F}_3$ for the spintensors
$f^{\lambda\mu}$ and $f^{\dot{\lambda}\dot{\mu}}$ (spinor representations
of the electromagnetic tensor), where $F_i=E_i-iB_i$ and
$\overset{\ast}{F}_i=E_i+iB_i$. Explicit expressions of the spinor
representations of the electromagnetic tensor are
\begin{gather}
\boldsymbol{\tau}_{1,0}:\quad\left\{\begin{array}{ccl}
f^{11}&\sim&4(F_1+iF_2),\\
f^{12}&\sim&4F_3,\\
f^{22}&\sim&4(F_1-iF_2);
\end{array}\right.\nonumber\\
\boldsymbol{\tau}_{0,1}:\quad\left\{\begin{array}{ccl}
f^{\dot{1}\dot{1}}&\sim&4(\overset{\ast}{F}_1+i\overset{\ast}{F}_2),\\
f^{\dot{1}\dot{2}}&\sim&4\overset{\ast}{F}_3,\\
f^{\dot{2}\dot{2}}&\sim&4(\overset{\ast}{F}_1-i\overset{\ast}{F}_2).
\end{array}\right.\nonumber
\end{gather}
In such a way, we see that complex linear combinations $\bF=\bE-i\bB$ and
$\overset{\ast}{F}=\bE+i\bB$, transformed within
$\boldsymbol{\tau}_{1,0}$ and $\boldsymbol{\tau}_{0,1}$ representations,
are coincide with the Majorana-Oppenheimer wave functions (\ref{MO1}) and
(\ref{MO2}). Therefore, the Majorana-Oppenheimer formulation of quantum
electrodynamics is a direct consequence of the group theoretical
framework of quantum field theory. In virtue of the Weinberg Theorem
\cite{Wein} for the Maxwell field $(1,0)\oplus(0,1)$
(or $(1,0)\cup(0,1)$) we have two helicity states $\lambda=1$ and
$\lambda=-1$ (left- and right-handed polarizations). In contrast to this,
the electromagnetic four-potential $A_\mu$ is described within
$\boldsymbol{\tau}_{\frac{1}{2},\frac{1}{2}}$-representation of the Lorentz
group. Therefore, the Gupta--Bleuler quantum electrodynamics,
based on the quantization of the non--observable quantity
$A_\mu$,\footnote{It should be noted here that at all times in
electrodynamics the four-potential $A_\mu$ is understood as an
auxiliary mathematical tool.} has led to the null helicity, what, as is known,
contradicts with experience. For that reason many authors
(see, for example, \cite{RF}) considered the Gupta-Bleuler quantization
of the electromagnetic field in terms of $A_\mu$ as a phenomenological
description. Moreover, such a description is not incorporated properly
into a group theoretical scheme of quantized fields.

In 1932, Majorana proposed the first construction of a relativistically
invariant theory of arbitrary half integer or integer spin particles
\cite{Maj32}. As is known \cite{Rec90}, the photon case is an initial point of
this construction. There is a close relationship between Majorana's
construction for the photon and other higher spin formalisms proposed
well after. For example, Weinberg considered the following
$(1,0)\oplus(0,1)$-field equations \cite{Wein}:
\begin{eqnarray}
\nabla\times[\bE-i\bB]+i(\partial/\partial t)[\bE-i\bB]&=&0,\nonumber\\
\nabla\times[\bE+i\bB]-i(\partial/\partial t)[\bE+i\bB]&=&0,\nonumber
\end{eqnarray}
which, as it easy to see, are equivalent to the equations (\ref{2}) and
(\ref{4}). At present time the Majorana-Oppenheimer formulation of
quantum electrodynamics is studied in terms of the
Joos-Weinberg formalism (with respect to Hammer-Tucker and Proca
equations) and the Bargmann-Wigner theory
\cite{Dvo97b}.

One of the most powerful higher spin formalisms is a Gel'fand-Yaglom
approach \cite{GY48} based primarily on the representation theory of
the Lorentz group. In contrast to the Bargmann-Wigner and
Joos-Weinberg formalisms, the main advantage of the Gel'fand-Yaglom
formalism lies in the fact that it admits naturally a Lagrangian
formulation. Indeed, an initial point of this theory is the following
lagrangian \cite{GY48,GMS,AB}:
\begin{equation}\label{Lag}
\cL=-\frac{1}{2}\left(\bar{\psi}\Gamma_\mu\frac{\partial\psi}{\partial x_\mu}-
\frac{\partial\bar{\psi}}{\partial x_\mu}\Gamma_\mu\psi\right)-
\kappa\bar{\psi}\psi,
\end{equation}
where $\Gamma_\mu$ are $n$--dimensional matrices, $n$ equals to the
number of components of the wave function $\psi$. Varying independently
the functions $\psi$ and $\bar{\psi}$, one gets general Dirac-like
(Gel'fand-Yaglom \cite{GY48}) equations
\begin{eqnarray}
\Gamma_\mu\frac{\partial\psi}{\partial x_\mu}+\kappa\psi&=&0,\nonumber\\
\Gamma^{\sT}_\mu\frac{\partial\bar{\psi}}{\partial x_\mu}-
\kappa\bar{\psi}&=&0.\nonumber
\end{eqnarray}
Let us consider a massless case of (\ref{Lag}):
\begin{equation}\label{Lag2}
\cL_M=-\frac{1}{2}\left(\boldsymbol{\psi}^\ast\alpha_\mu
\frac{\partial\boldsymbol{\psi}}{\partial x_\mu}-
\frac{\partial\boldsymbol{\psi}^\star}{\partial x_\mu}\alpha_\mu
\boldsymbol{\psi}\right),
\end{equation}
where $\boldsymbol{\psi}=\bE-i\bB$, $\boldsymbol{\psi}^\ast=\bE+i\bB$, and
$\alpha_\mu$ ($\mu=0,1,2,3$) are the matrices (\ref{3}), at this point
$\alpha_0=W$ is a unit matrix. Varying independently the functions
$\boldsymbol{\psi}$ and $\boldsymbol{\psi}^\ast$ in (\ref{Lag2}), we obtain
\begin{eqnarray}
\frac{\partial\cL}{\partial\boldsymbol{\psi}^\ast}=
-\frac{1}{2}\alpha_\mu\frac{\partial\boldsymbol{\psi}}{\partial x_\mu},\quad
\frac{\partial\cL}{\partial\frac{\partial\boldsymbol{\psi}^ast}
{\partial x_\mu}}=\frac{1}{2}\alpha_\mu\boldsymbol{\psi};\label{V1}\\
\frac{\partial\cL}{\partial\boldsymbol{\psi}}=
\frac{1}{2}\frac{\partial\boldsymbol{\psi}^\ast}
{\partial x_\mu}\alpha_\mu,\quad
\frac{\partial\cL}{\partial\frac{\partial\boldsymbol{\psi}}
{\partial x_\mu}}=-\frac{1}{2}\boldsymbol{\psi}\alpha_\mu.\label{V2}
\end{eqnarray}
Substituting the relations (\ref{V1}) and (\ref{V2}) into the Euler equation
\[
\frac{\partial\cL}{\partial u_i}-\frac{\partial}{\partial x}
\frac{\partial\cL}{\partial\frac{\partial u_i}{\partial x}}=0,
\]
we come to equations
\begin{eqnarray}
\alpha_\mu\frac{\partial\boldsymbol{\psi}}{\partial x_\mu}&=&0,\nonumber\\
\alpha^{\sT}_\mu\frac{\partial\boldsymbol{\psi}^\ast}{\partial x_\mu}&=&0
\nonumber
\end{eqnarray}
which, obviously, are equivalent to the equations (\ref{2}) and (\ref{4}).
At this point, we have $j_\mu\neq 0$ (so-called neutral current), and
the `charge' $Q=\int d\bx j_0(x)=\int d\bx\boldsymbol{\psi}^\ast\alpha_0
\boldsymbol{\psi}$ is proportional to the energy $\bE^2+\bB^2$ of
electromagnetic field.

In conclusion it should be noted that Majorana-Oppenheimer
formulation is an antithesis to the widely accepted gauge paradigm based
on the so-called Standard Model, in which all the physical fields
are divided into the `gauge' and `matter' fields. One of the main
preferences of the Majorana-Oppenheimer quantum electrodynamics lies in the
fact that it allows one to avoid this opposition and to consider all the
fields in equal footing. Moreover, the Majorana-Oppenheimer formulation
based naturally on the ground of group theoretical methods which include
a wide variety of powerful mathematical tools.

\end{document}